# The Electromagnetic Waves Propagation in Unmagnetized Plasma Media Using Parallelized Finite-Difference Time-Domain Method

Lang-lang Xiong [1], Xi-min Wang [1], Song Liu[*, 1, 2], Zhi-yun Peng [1], and Shuang-ying Zhong [1]

1．Nanchang University, Nanchang Jiangxi 330031, P. R. China.
2．State Key Laboratory of Millimeter Waves, Nanjing Jiangsu 210096, P. R. China.
* Corresponding author E-mail address: sliu@ncu.edu.cn

*Abstract*—The finite-difference time-domain (FDTD) method has been commonly utilized to simulate the electromagnetic (EM) waves propagation in the plasma media. However, the FDTD method may bring about extra run-time on concerning computationally large and complicated EM problems. Fortunately, the FDTD method is easy to parallelize. Besides, GPU has been widely used for parallel computing due to its unique SPMD (Single Program Multiple Data) architecture. In this paper, we represent the parallel Runge-Kutta exponential time differencing scheme FDTD (RKETD) method for the unmagnetized plasma implemented on GPU. The detailed flowchart of parallel RKETD-FDTD method is described. The accuracy and acceleration performance of the proposed parallel RKETD-FDTD method implemented on GPU are substantiated by calculating the reflection and transmission coefficients for one-dimensional unmagnetized plasma slab. The results indicate that the numerical precision of the parallel RKETD-FDTD scheme is consistent with that of the code implemented on CPU. The computation efficiency is greatly improved compared with merely CPU-based serial RKETD-FDTD method. Moreover, the comparisons of the performance of CUDA-based GPU parallel program, OpenMP (Open Multi-Processing)-based CPU parallel program, and single-CPU serial program on the same host computer are done. Compared with the serial program, both parallel programs get good results, while GPU-based parallel program gains better result.

## I. INTRODUCTION

Since the finite-difference time-domain (FDTD) method was initially delivered to numerically resolve the Maxwell's equations by Yee in 1966 [1]. It has been widely used in the numerical solution of electromagnetics (EMs) problems. The FDTD method has obvious advantages compared to many other numerical methods. It uses the leap frog algorithm—the electric field and the magnetic field in the space domain to perform alternate calculations. So do not need too complicated calculations[2].

Over the past decades, the FDTD numerical modeling approach has been applied to many aspects, including the modeling of objects in aerospace, biological systems and geometric shapes, the analyzing and designing of complicated microwave circuits, fast time-varying systems and other engineering applications [3]. Plentiful numerical methods related to FDTD formulations used to calculate the EM waves propagation in the dispersive media are addressed, including the recursive convolution (RC) method [4,5], frequency-dependent Z transform method [6,7], direct integration (DI) method [8, 9], JE convolution (JEC) method [10], the auxiliary differential equation (ADE) method [11], piecewise linear recursive convolution (PLRC) method [12], piecewise linear current density recursive convolution (PLCDRC) method [13], and Runge-Kutta exponential time differencing (RKETD) method [14]. Simulation of EM waves propagating through the plasma media is a unique and fascinating application built on FDTD formulations for dispersive media. The appearing nonlinear phenomena that are not totally understood can be explicitly refined by numerical simulation. Furthermore, the aforementioned various FDTD scheme for dispersive media can be applied to the plasmas.

Although the FDTD schemes above are well-suited to numerical simulation, the original FDTD method can bring about extra run time owing to computationally large and complicated EM issues. However, FDTD method is naturally a massively parallel algorithm, thus it can benefit a lot from the progresses in parallel computing techniques and effectively reduce the run time. In 2001, an MPI-based parallel FDTD algorithm was proposed [15]. And some physicists used VALU and VAX to speed up parallel FDTD procedures [16]. However, the acceleration of FDTD algorithm was not significant and not easy to implement. OpenMP-based CPU parallel program can be easily achieved in a single host. Researchers have applied it to the FDTD algorithm [17][18], but the acceleration performance has not been improved greatly. And it is difficult to implement parallel computation on multiple hosts. More and more researches begin to focus on the parallelization of FDTD algorithm by GPU to improve computational efficiency. Patrick D. Cannon and Farideh Honary utilized OpenCL-based GPU to implement parallel FDTD program and successfully simulated electromagnetic wave propagation in plasma [19]. OpenCL, however, was unfriendly to developers and it did not occupy the mainstream market of general parallel computing. Wang et al, used NVIDIA GPU based CUDA(Compute Unified Device Architecture) parallelization FDTD method and calculated the electromagnetic field propagation in the dielectric in 2016[20]. In the same year Ryo Imai et al compared the FDTD performance under PML boundary conditions using GPU, MIC and CPU[21]. In 2017, Diener and others compared the CUDA program and MATLAB parallel computing toolbox parallel FDTD method acceleration performance. And numerical simulation results show that CUDA algorithm are more efficient [22]. What's more, FDTD method carried out on multi-GPU clusters has triggered great interest to further accelerate large-scale computations with improved speedup performance [23-26]. However, for complicated medium, dispersive media for instance, the research of parallel FDTD algorithm based on CUDA platform has not been used. So, a parallel FDTD method with higher accuracy and efficiency capable of computing complicated medium is proposed in this paper.

This paper presents a GPU-based parallel RKETD-FDTD method with CUDA for the unmagnetized plasma media to acquire better acceleration performance, compared with merely CPU-based serial RKETD-FDTD method. Numerical simulation of the RKETD-FDTD method for the unmagnetized plasma media is undertaken both on CPU and GPU respectively. The reflection and transmission coefficients through an unmagnetized plasma layer in one dimension are calculated to validate the accuracy of the method. Comparing the CPU-based serial program with the CPU-based parallel program, the calculated acceleration ratio proves the high efficiency of GPU-based parallel RKETD-FDTD method.

This paper is arranged as follows. In Section II, we describe the Maxwell equations for the unmagnetized plasma and derive the FDTD formulation with RKETD numerical scheme. Section III illustrates the implementation of GPU-parallelized RKETD-FDTD method with CUDA. Section IV designs a numerical simulation that is carried out to prove the accuracy and the efficiency of the GPU-based parallel RKETD-FDTD method for unmagnetized plasma.

## II. RKETD-FDTD Formulation

The famous Maxwell's equations in time domain for the unmagnetized plasma are provided by

$$\nabla \times \boldsymbol{H} = \varepsilon_0 \frac{\partial \boldsymbol{E}}{\partial t} + \boldsymbol{J} \tag{1}$$

$$\nabla \times \boldsymbol{E} = -\mu_0 \frac{\partial \boldsymbol{H}}{\partial t} \tag{2}$$

$$\frac{d\boldsymbol{J}}{dt} + \nu \boldsymbol{J} = \varepsilon_0 \omega_p^2 \boldsymbol{E} \tag{3}$$

where, $\boldsymbol{H}$ is the magnetic intensity, $\boldsymbol{E}$ is the electric field, $\boldsymbol{J}$ is the polarization current density, $\nu$ is the electron collision frequency, $\omega_p$ is plasma frequency, $\varepsilon_0$ and $\mu_0$ are the permittivity and permeability of free space, respectively. Considering one-dimensional equations, the one-dimensional component of the equations can be written as

$$-\frac{\partial H_y}{\partial z} = \varepsilon_0 \frac{\partial E_x}{\partial t} + J_x \tag{4}$$

$$\frac{\partial E_x}{\partial z} = -\mu_0 \frac{\partial H_y}{\partial t} \tag{5}$$

$$\frac{dJ_x}{dt} + \nu J_x = \varepsilon_0 \omega_p^2 E_x \tag{6}$$

The Runge-Kutta exponential time differencing (RKETD) scheme [27] is derived in detail.

Multiplying (6) through by the integrating factor $e^{\nu t}$ and integrating the equation over a single time step, specifying $t_{n+1} = t_n + \Delta t$, $t = t_n + \tau$. The exact solution from $t_n$ to $t_{n+1}$ is

$$J_x(t_{n+1}) = e^{-\nu \Delta t} J_x(t_n) + e^{-\nu \Delta t} \int_0^{\Delta t} e^{\nu \tau} F(J_x(t_n + \tau), t_n + \tau) d\tau \tag{7}$$

where $F(J_x(t_n + \tau), t_n + \tau) = \varepsilon_0 \omega_p^2 E_x(t_n + \tau)$. While the right term $\int_0^{\Delta t} e^{\nu \tau} d\tau$ is in deriving approximations to the integral in eq.(7), $J_x^{n+1}$ denotes the numerical approximations to $J_x(t_{n+1})$, and $F(J_x(t_n), t_n) = F_n$. The simplest approximation to the integral is that $F$ is constant, namely

$$F = F_n + O(\Delta t) \tag{8}$$

The numerical solution of eq.(7) is written as

$$J_x^{n+1} = J_x^n e^{-\nu \Delta t} - F_n (e^{-\nu \Delta t} - 1)/\nu. \tag{9}$$

The numerical accuracy of the above method (8) is only first-order. If $F$ changes over the interval $t_n \leq t \leq t_{n+1}$, the second-order approximation of ETD method can be expressed as

$$F = F_n + \frac{\tau}{\Delta t}[F_n - F_{n-1}] + o((\Delta t)^2). \tag{10}$$

Substitution of equation (10) in equation (7), after some manipulation the x component of $J$ at $n+1$ time step can be written as

$$J_x^{n+1} = e^{-\nu \Delta t} J_x^n + \frac{(1 - e^{-\nu \Delta t})}{\nu} \varepsilon_0 \omega_p^2 E_x^n + \frac{(e^{-\nu \Delta t} - 1 + \nu \Delta t)}{\nu^2 \Delta t}[\varepsilon_0 \omega_p^2 E_x^{n+1} - \varepsilon_0 \omega_p^2 E_x^n]. \tag{11}$$

So equation (11) can be used to update the x component of the polarization current density $J$ by utilizing the x component of the electric field $E$ at both $n$ and $n+1$ time steps.

The discretization of equation (4) follows the Yee grid and leapfrog-style algorithm, which is taken to give

$$E_x^{n+1}(k) = E_x^n(k) - \frac{\Delta t}{\varepsilon_0 \Delta z}\left[H_y^{n+\frac{1}{2}}(k+\frac{1}{2}) - H_y^{n+\frac{1}{2}}(k-\frac{1}{2})\right] - \frac{\Delta t}{2\varepsilon_0}\left[J_x^{n+1}(k) + J_x^n(k)\right] \tag{12}$$

Substituting equation (11) into equation (12) gives

$$E_x^{n+1}(k) = (1 + \frac{\frac{\omega_p^2 \Delta t}{2\nu}(e^{-\nu \Delta t} - 1)}{1 + \frac{\omega_p^2}{2\nu^2}(e^{-\nu \Delta t} - 1 + \nu \Delta t)}) E_x^n(k) - \frac{\Delta t}{2\varepsilon_0}(1 + e^{-\nu \Delta t}) J_x^n(k) - \frac{\Delta t}{\varepsilon_0 \Delta z}[H_y^{n+\frac{1}{2}}(k+\frac{1}{2}) - H_y^{n+\frac{1}{2}}(k-\frac{1}{2})] \tag{13}$$

The discretization of equation (5) strongly resembles the FDTD formulation for vacuum, and the update equation of $H$ is given by

$$H_y^{n+\frac{1}{2}}(k+\frac{1}{2}) = H_y^{n-\frac{1}{2}}(k+\frac{1}{2}) - \frac{\Delta t}{\mu_0 \Delta x}[E_x^n(k+1) - E_x^n(k)] \tag{14}$$

III. THE GPU-PARALLELIZED FDTD IMPLEMENTATION

In the FDTD method, the iteration of any field needs its adjacent field. Therefore, when the iteration of the field at the boundary of the sub-region is performed in parallel, the field information needs to be transmitted. However, information is needed to be transmitted only between adjacent fields, and the value of each sub-area is not required to transmit. Consequently, FDTD method is very suitable for parallel computing.

For parallel computing, SPMD architecture[28] has been put forward. The SPMD parallel computer is composed of multiple computers or processors of the same status. Only one program runs at the same time. The synchronization between different processors is ensured by the developers. Those processes proved that SPMD architecture is suitable for parallel algorithms. The calculation space is divided into several blocks, each on a processor, but all processors perform the same program. CUDA-based GPU is the use of SPMD architecture. The GPU has far more processors than the CPU, and each processor in the GPU is made up of multiple threads, all of which can execute the same program in parallel. More transistors are used for data processing than for data caching and process control over CPUs. However, because the same program is executing on each element, there are very few requirements for complex process control. And because multiple elements execute with high computational density, the fetch latency can be computationally hidden, thus eliminating the need for large data caches.

Compared with RKETD-FDTD method on CPU, we illustrate the flowchart of GPU-parallelized RKETD-FDTD method displayed in Figure 1. In the actual program, we divide the code into the host code running on the CPU and the device code running on the GPU. The host code is implemented to complete fields and parameters initialization, device memory allocation and release, and data transfer between host and device. To efficiently implement GPU-parallelized RKETD-FDTD method with CUDA, device codes are considered to be divided into two kernels: H-update-kernel to update the magnetic intensity and E-update-kernel to update the electric field. Because of the use of PML boundary conditions, both the electric and magnetic fields are divided into two directions, both of which execute in their own kernel.

The two kernels with dashed lines are launched sequentially and executed in parallel on GPU. High-speed exchange of data between kernels is achieved through bandwidth-intensive shared memory. The execution configuration for each kernel is governed by the complete domain and allocated shared memory, for the reason that the calculating of a Yee's cell is mapped to a thread on the device.

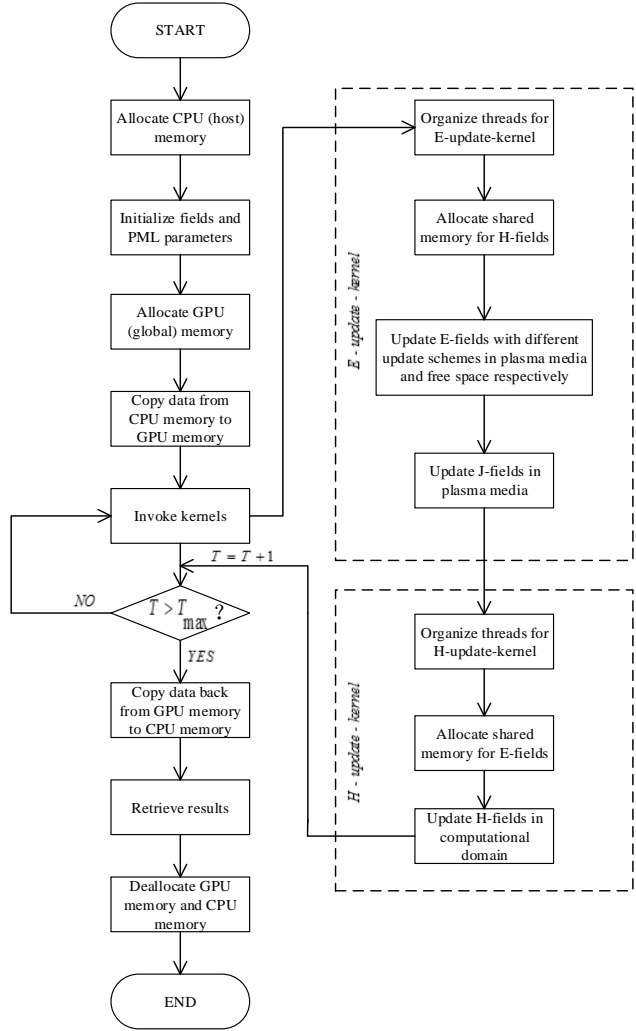

Figure 1 Flowchart of GPU-parallelized FDTD method with CUDA

## IV. NUMERICAL SIMULATION

*A. Simulation environment*

As briefly stated before, it's theoretically predicted that the GPU-based parallel RKETD-FDTD method can acquire better acceleration performance compared with the corresponding simply CPU-based serial RKETD-FDTD method, when RKETD-FDTD method is utilized to simulate EM waves traveling through the unmagnetized plasma media. Nevertheless, the application acceleration performance of the GPU-RKETD-FDTD method isn't always as satisfying as estimated in practical applications, which frequently varies with problem complexities, GPU hardware properties and programmers' skills.

To evaluate the acceleration performance of GPU-RKETD-FDTD method, we prefer CUDA as the parallel programming platform to employ the acceleration of the GPU-RKETD-FDTD method for the unmagnetized plasma media. The numerical experiment is performed on the computer provided with CUDA-supported NVIDIA GPU. The CPU is Intel Core i7 6700 with a 3.4GHz processor. It has four cores, eight threads with 16GB of host memory. And the GPU is NVIDIA GeForce GT 1080. GeForce GT 1080 is based on graphics processor of NVIDIA Pascal architecture, and is developed for desktop applications. Some key specifications of GT1080 are tabulated in Table 1. The development environment is Microsoft Visual Studio 2015 (Community Edition) with CUDA toolkit 9.1 assembled, Windows 10 as operating system.

Table 1 Specifications of NVIDIA GeForce GT 1080

| Specification | GeForce GT 1080 |
| --- | --- |
| Chip | GP104-400 |
| CUDA cores | 2560 |
| Processor clock | 1607MHz |
| Memory clock | 10010MHz |
| Memory size | 8192MB |

B. *Numerical Results*

To exhibit the simulation of above mentioned RKETD-FDTD formulation for the unmagnetized plasma media, we choose appropriate parameters that are listed as follows. The entire computational domain is $4096\Delta z$ in $z$ axis, where $\Delta z$ is selected to be $\Delta z = 75\mu m$ as size of a cell. A single time step is taken to be $\Delta t = 125 fs$. The unmagnetized plasma slab with the thickness of $9cm$ occupies the middle 1200 cells. The PML absorbing boundary conditions [29] are implemented at both ends of 5 cells to avoid undesired reflections, and the rest space is vacuum. The key parameters of unmagnetized plasma are $w_p = 2\pi \times 28.7 \times 10^9 rad/s$ and $v = 20GHz$. The source wave exploited in numerical experimentation is the Gaussian-derivative pulsed plane wave, which is given in time domain by

$$E_i(t) = (t - 5\tau)\exp\left(-\frac{(t-5\tau)^2}{2\tau^2}\right), \quad (15)$$

here, $\tau = 15\Delta t$.

The execution configurations of the two kernels are arranged both as 16 blocks in each grid and 256 threads in each block in case of the computational domain, with CUDA programming model constructed. To validate the high accuracy of the GPU-RKETD-FDTD method with CUDA compared with the accuracy using RKETD-FDTD method merely CPU-based, reflection and transmission coefficients are analyzed with the Fast Fourier Transform (FFT). Figure 2 and 3 clarify the magnitudes of reflection coefficient and transmission coefficient computed respectively by GPU-RKETD-FDTD method on GPU and RKETD-FDTD method on CPU with those of the analytical solution.

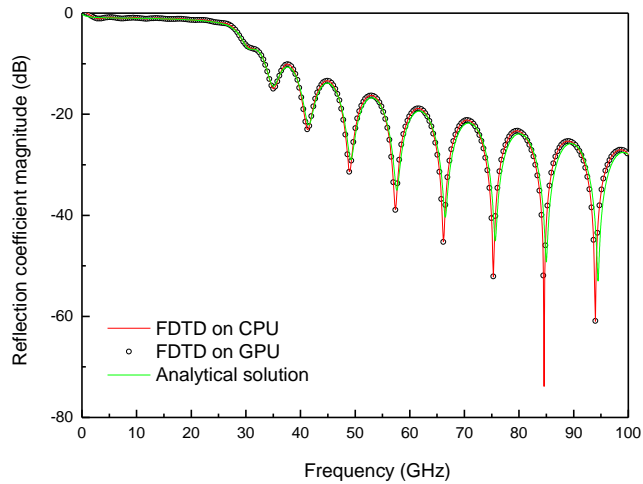

Figure 2 (color online) Reflection coefficient magnitude versus frequency

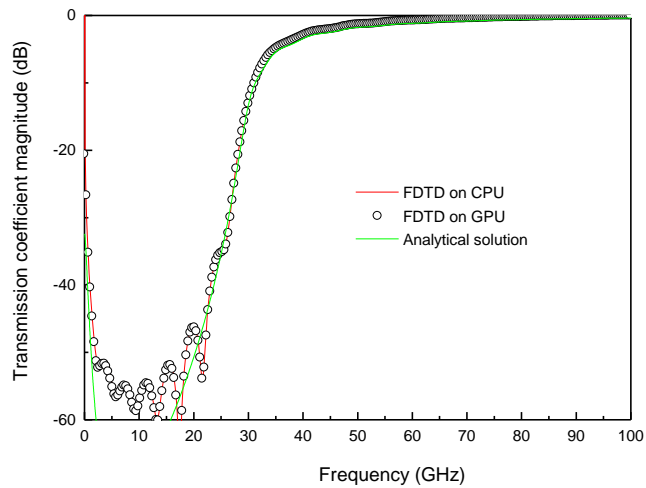

Figure 3 (color online) Transmission coefficient magnitude versus frequency

Table 2 illustrates the run time of GPU-based parallel programs, CPU-based parallel programs and CPU serial programs, and their corresponding speed-up ratios. The speed-up ratio is obtained by dividing serial program operation time by parallel program operation time under the same Yee's Cells and time step. Speedup ratios I and II refer to the ratios of GPU parallel programs and CPU parallel programs, CPU serial programs respectively. To calculate the efficiency of different methods when Yee's Cells increase, we keep the time step invariant at 1000 steps.

The speedup ratio I is increasing with time step increasing. When the cells are less than 20,480, the calculation time of GPU parallel FDTD program does not change much. This is because the GPU is not fully utilized when the cells are small.

The speed-up ratio in Table 2 shows that the GPU-RKETD-FDTD method is more efficient than the CPU serial program and the OpenMP-based CPU parallel program.

Table 2 Speedup ratio at diverse Yee's cells in the simulation

| Yee's Cells | GPU Times (ms) | CPUs Times(ms) | CPU Times (ms) | Speedup Ratios I | Speedup Ratios II |
|---|---|---|---|---|---|
| 1,024 | 33 | 114 | 279 | 8.45 | 2.45 |
| 5,120 | 33 | 122 | 379 | 11.48 | 3.11 |
| 10,240 | 33 | 138 | 504 | 15.27 | 3.65 |
| 20,480 | 34 | 148 | 753 | 22.15 | 5.08 |
| 30,720 | 46 | 175 | 982 | 21.35 | 5.61 |
| 40,960 | 53 | 231 | 1,231 | 23.23 | 5.33 |
| 51,200 | 61 | 299 | 1,467 | 24.05 | 4.91 |
| 61,440 | 64 | 309 | 1,701 | 26.58 | 5.50 |

We can easily make further improvement in speedup performance for practical applications when the numerical results of GPU-RKETD-FDTD method are generalized to the two-dimensional and even three-dimensional conditions. Compared with the one-dimensional condition, it can make full use of the GPU computing resources, instead of confirming the speedup performance only. Furthermore, more sophisticated CUDA-supported GPU, such as Tesla, can be upgraded to obtain better speedup performance.

## V. CONCLUSION AND PERSPECTIVES

In this letter, we presented a CUDA-based parallel RKETD-FDTD method for the unmagnetized plasma implemented on GPU to acquire better acceleration performance, compared with the previous RKETD-FDTD method simply implemented on single CPU or multiple CPUs. Numerical experiments of these methods have been done respectively on NVIDIA GPU and on CPU. The accuracy and acceleration performance of the represented GPU-RKETD-FDTD method were assessed by the numerical simulation of the EM waves traveling through the unmagnetized plasma slab, compared with CPU-based RKETD-FDTD method. The accuracy was verified by calculating the reflection and transmission coefficients for one-dimensional unmagnetized plasma slab. The comparison between the elapsed times of these methods has proved that the proposed GPU-RKETD-FDTD method implemented with CUDA can acquire decent acceleration with sufficient accuracy. Further research will be predicted to obtain more satisfactory acceleration performance while the numerical results of GPU-RKETD-FDTD method are generalized to two-dimensional and even three-dimensional conditions for practical applications. Besides, more sophisticated CUDA-supported GPUs specialized in scientific computation such as Tesla can be utilized to acquire better acceleration performance.

**Acknowledgment**

This work was supported by National Nature Science foundation of China (No. 61261006, 11165011 & 11563006 ), the State Key Laboratory of Millimeter Waves Open Research Program (No. K201605). Our deepest gratitude goes to the anonymous reviewers and editors for their careful work and thoughtful suggestions that have helped improve this paper substantially. Thanks are due to Zhibin Xie for assistance with the experiments and to Luo Li, Cen Zhang, Huaqin Sun and Sujia Chen for valuable discussion.